\documentclass[10pt,prl,twocolumn]{revtex4}%
\usepackage{amsfonts}
\usepackage{amsmath}
\usepackage{amssymb}
\usepackage{graphicx}
\usepackage[usenames]{color}%
\providecommand{\U}[1]{\protect\rule{.1in}{.1in}}
\begin{document}
\title{Shear viscosity coefficient and relaxation time of causal dissipative
hydrodynamics in QCD}
\author{T. Koide}
\affiliation{FIAS, Johann Wolfgang Goethe-University\"at, Ruth-Moufang Str. 1, 60438,
Frankfurt am Main, Germany}
\author{E. Nakano}
\affiliation{Extreme Matter Institute, GSI, Planckstr. 1, D-64291 Darmstadt, Germany}
\author{T. Kodama}
\affiliation{Instituto de F\'{\i}sica, Universidade Federal do Rio de Janeiro, C. P. 68528,
21945-970, Rio de Janeiro, Brasil}

\begin{abstract}
The shear viscosity coefficient and the corresponding relaxation time for
causal dissipative hydrodynamics are calculated based on the microscopic
formula proposed in [T. Koide and T. Kodama, Phys. Rev. \textbf{E 78}, 051107
(2008)]. Here, the exact formula is transformed into a more compact form and
applied it to evaluate these transport coefficients in the chiral perturbation
theory and perturbative QCD. It is shown that in the leading order
calculation, the causal shear viscosity coefficient $\eta$ reduces to that of
the ordinary Green-Kubo-Nakano formula, and the relaxation time $\tau_{\pi}$
is related to $\eta$ and pressure $P$ by a simple relationship, $\tau_{\pi
}=\eta/P$.

\end{abstract}
\maketitle

The study of dissipative processes in relativistic heavy ion collisions is now
one of the important topics to clarify quantitatively how closely the matter
created there behaves as an ideal fluid. The proper concept of relativistic
dissipative hydrodynamics is, however, not trivial at all. It is by now known
that a naive covariant extension of the Navier-Stokes equation leads to the
problem of relativistic acausality and instabilities \cite{dkkm3}.

An essential key to this question is the presence of memory effect
characterized by a finite relaxation time in the definition of irreversible
currents \cite{dkkm4}. Consequently, in a relativistic regime, any fluid
becomes \textit{non-Newtonian}, that is, the irreversible current is not
simply proportional to thermodynamic forces. 
We refer to those
theories which incorporate this effect as the causal
dissipative hydrodynamics (CDH).

Similarly to equation of states, transport coefficients reflect the properties
of the matter. These are inputs for hydrodynamic modelings and should be
determined from a microscopic theory. So far, there are mainly two approaches
to estimate these coefficients of CDH. One is the kinetic approach
based on the Boltzmann equation \cite{muronga,is}. The other is the
calculation from a duality assuming the AdS/CFT (Anti-de Sitter space/
conformal field theory) correspondence \cite{son,baier}. However, the
applicability of these approaches is not obvious for the physics of
heavy-ion collisions. The kinetic approach is only applicable for rarefied
gases where the Boltzmann-Grad limit is satisfied. The AdS/CFT approach
predicts the behavior of a matter described by a conformal field theory in its
the strong coupling limit, but QCD is obviously not a conformal theory.

The Green-Kubo-Nakano (GKN) formula is another possibility which is used to
evaluate these transport coefficients in terms of microscopic correlation
functions.
%Some calculations
%are found in the lattice QCD approach \cite{naka}.
However, for our purpose, there exist the following problems. One is that, in
this approach, the relaxation time has not been expressed with correlation
functions. Furthermore, it is not obvious whether the GKN formula for the
shear viscosity coefficient is applicable when the system exhibits a
non-Newtonian nature of irreversible currents. Usually Newtonian nature is
assumed in the derivation of the GKN formula but, as mentioned above, this is
exactly the key question for the case of a relativistic fluid \cite{koide,kk}.
More precisely, the shear viscosity coefficient appears as a phenomenological
parameter, that is, the diffusion constant for transverse momentum via a ratio
to entropy density $s$ in the phenomenological Langevin approach. However,
this proportionality to the entropy density is not always true and depends on
its definition from a microscopic theory.

Another important point is that the hydrodynamic shear viscosity, for example,
appears in a thermal relaxation process which is not a response of the system
to an external force, but is rather a response to the inhomogeneity of the
velocity field, that is, the change in boundary conditions. Thus a direct
application of the traditional argument of the linear response theory for the
calculation of the shear viscosity is not straightforward. As a matter of
fact, the GKN formula of the shear viscosity is derived by using, for example,
the nonequilibrium statistical operator method \cite{hosoya}.

Recently, the microscopic formulae to calculate the transport coefficients of
CDH was derived by use of the projection operator method
\cite{koide,kk,b-zwanzig}. Although this approach is shown to be very
powerful, the results obtained so far are yet too formal for the practical
applications. The aim of the present work is two fold: one is to reduce the
formula obtained in \cite{koide,kk} into a more compact form, and the other is
to apply it to the hadronic matter and quark-gluon plasma (QGP) to investigate
the temperature dependence of the causal shear viscosity coefficient $\eta$
and the relaxation time $\tau_{\pi}$.

So far, $\tau_{\pi}$ has not been discussed enough compared to $\eta$, 
but some crucial questions depend on the precise
behavior of the relaxation time. For example, the behavior of $\tau_{\pi}$ is
directly related to the stability of relativistic fluids as discussed in
\cite{dkkm3}. Another question is the value of $\eta/s$ obtained form the
analysis of experimental data seems very small, as close, if not smaller than,
to the lower bound $1/4\pi$ predicted by the AdS/CFT correspondence. However,
even if $\eta/s$ is not so small, the fluid can behave as an ideal fluid when
$\tau_{\pi}$ is sufficiently large. Thus to know the precise value of
$\tau_{\pi}$ is fundamental.

For simplicity, let us consider a system with shear flow, where the fluid
velocity points to the $x$ direction with finite velocity gradient in the $y$
direction. We further assume that the evolution equation of the shear
viscosity is approximately given by the evolution of the two variables,
$T^{yx}$ and $T^{0x}$. Correspondingly, we will employ the formula of $N=2$
form given in Sec.~VIII of \cite{kk}, where $N$ is the dimension of the
projected space. The exact time evolution of spatial, off-diagonal components
of energy-momentum tensor $T^{\mu\nu}(\mathbf{k},t)$ in the momentum
representation (Fourier transform) is formally expressed as
\begin{align}
\partial_{t}T^{yx}(k_{y},t)  &  =-ik_{y}R_{k_{y}}T^{0x}(k_{y},t)\nonumber\\
&  \hspace{-1cm}-\int_{0}^{t}d\tau\Xi_{22}(k_{y},\tau)T^{yx}(k_{y},t-\tau
)+\xi_{k_{y}}(t), \label{eq_exact2}%
\end{align}
where $R_{k_{y}}=(T^{yx}(k_{y}),T^{yx}(-k_{y}))/(T^{0x}(k_{y}),T^{0x}%
(-k_{y})),$ and the inner product $(X,Y)$ represents Kubo's canonical
correlation $(X,Y)=\int_{0}^{\beta}d\lambda\beta^{-1}\mathrm{Tr}[\rho
_{eq}e^{\lambda H}Xe^{-\lambda H}Y]$. Here, $H$ is the Hamiltonian and $\beta$
is the inverse of temperature $1/T$. The memory function $\Xi_{22}(k_{y},t)$
is defined in Eq. (78) of \cite{kk}. The last term of the Eq. (\ref{eq_exact2}%
) is often called the noise term, and related to the memory term through the
fluctuation-dissipation theorem of the second kind \cite{koide,kk}.

To break the time-reversal symmetry, we introduce the time-convolutionless
(TCL) approximation\cite{koide,kk},
\begin{equation}
\partial_{t}T^{yx}(k_{y},t)=-ik_{y}(\varepsilon+P)R_{k_{y}}u^{x}%
(k_{y},t)-\frac{1}{\tau_{k_{y}}}T^{yx}(k_{y},t),\label{eq_markov2}%
\end{equation}
where we have introduced a function related to the relaxation time
$\tau_{k_{y}}^{-1}=\int_{0}^{\infty}d\tau\Xi_{22}(k_{y},\tau)$, and
$\varepsilon$, $P$ and $u$ are the energy density, pressure and fluid
velocity, respectively. The noise term is not necessary for the present
discussion and dropped out. The TCL approximation is equivalent to assume the
exponential ansatz for the memory of $T^{yx}(k_{y},t)$. It should be
emphasized that, even after the TCL approximation, the hysteresis of $T^{xy}$
is still reserved in its time evolution equation and this corresponds exactly
to the memory effect in the usual CDH\cite{dkkm4}. In this derivation, we have
used the following replacement $T^{0x}=(\varepsilon+P)u^{x}(k_{y},t)$, which
is justified near the local rest frame.

On the other hand, the linearized phenomenological equation for the shear
viscosity defined by projection to the traceless part, 
$\pi^{\mu\nu}=\frac{1}{2}\left(  \Delta^{\mu\alpha}\Delta^{\nu\beta}
+\Delta^{\mu\beta}\Delta^{\nu\alpha}-\frac{2}{3}\Delta^{\mu\nu}\Delta^{\alpha\beta}\right)
T_{\alpha\beta}$ with $\Delta^{\mu\nu}=g^{\mu\nu}-u^{\mu}u^{\nu}$, is given
by
\begin{equation}
\tau_{\pi}\frac{\partial}{\partial t}\pi^{yx}(k_{y})+\pi^{yx}(k_{y}%
)=-\eta(ik_{y})u^{x}(k_{y}) \label{pheno}%
\end{equation}
at the local rest frame. Comparing the above equation with
Eq.~(\ref{eq_markov2}), we identify the $\eta$ and $\tau_{\pi}$ \cite{kk}:
\begin{equation}
\tau_{\pi}=\lim_{k_{y}\rightarrow0}\tau_{k_{y}},~~~\eta=\lim_{k_{y}%
\rightarrow0}(\varepsilon+P)R_{k_{y}}\tau_{k_{y}}. \label{de_shear}%
\end{equation}

The memory function can be cast into a simple form in the low momentum limit:
\begin{equation}
\Xi^{L}_{22}(\mathbf{k},s)=\frac{1-sX_{22}^{L}(k_{y},s)}{X_{22}^{L}(k_{y}%
,s)}+O(k_{y}),
\end{equation}
where $X_{22}^{L}(k_{y},s)$ is the Laplace transform with respect to $t$ of
the following correlation function,
\begin{equation}
X_{22}(k_{y},t)=\frac{(T^{yx}(k_{y},t),T^{yx}(-k_{y},0))}{(T^{yx}%
(k_{y},0),T^{yx}(-k_{y},0))}.
\end{equation}
Substituting this equation into the definitions (\ref{de_shear}), we obtain
$\tau_{\pi}=X_{22}^{L}(\mathbf{0},0)$ and $\eta=(\varepsilon+P)R_{\mathbf{0}%
}X_{22}^{L}(\mathbf{0},0)$.

For later convenience, we re-express the correlation function $X^{L}%
_{22}(\mathbf{0},0)$ with the definition of the GKN shear viscosity
coefficient \cite{hosoya} as
\begin{equation}
X^{L}_{22}(\mathbf{0},0) = \frac{\eta_{GKN}}{\beta}\left[ \int d^{3}\mathbf{x}
(\pi^{xy}(\mathbf{x},0),\pi^{xy}(\mathbf{0},0))\right] ^{-1},
\end{equation}
where
\begin{align}
\eta_{GKN} = -\frac{1}{10}\int d^{3} \mathbf{x} \int^{t}_{-\infty}dt_{1}
\int^{t_{1}}_{-\infty} d\tau\langle\pi^{\alpha\beta}(\mathbf{x},t) \pi
_{\alpha\beta}(\mathbf{0},\tau) \rangle_{ret}\nonumber\\
&  \hspace{-4cm}(\alpha,\beta= 0,1,2,3).
\end{align}
Here the suffix $ret$ denotes the retarded Green function, and there is
difference by factor 2 from \cite{hosoya} because of the difference of the definition.

Finally, $\eta$ and $\tau_{\pi}$ are given by
\begin{align}
\frac{\eta}{\beta(\varepsilon+P)}  &  = \frac{\eta_{GKN}}{\beta^{2} \int
d^{3}\mathbf{x} (J^{x}(\mathbf{x},0),J^{x}(\mathbf{0},0))},\label{csvc}\\
\frac{\tau_{\pi}}{\beta}  &  = \frac{\eta_{GKN}}{\beta^{2}\int d^{3}
\mathbf{x} (\pi^{xy}(\mathbf{x},0),\pi^{xy}(\mathbf{0},0))}, \label{relaxt}%
\end{align}
where we have introduced the energy current, $J^{\mu}(\mathbf{x},t)=u_{\nu
}T^{\nu\mu}(\mathbf{x},t)$ which is $T^{0\mu}(\mathbf{x},t)$ at the local rest
frame. Note that $\eta/(\beta(\varepsilon+P))$ is nothing but $\eta/s$ at
vanishing chemical potential. These expressions are still most general ones
within a framework where the hydrodynamic description is meaningful and are the
first main results of the present work.

Concerning the causality problem in CDH, we introduce a ratio
$\eta/(\tau_{\pi}(\varepsilon+P))$, which is, in our case, given by
$R_{\mathbf{0}}$ itself. Note that the propagation speed of signal should be
slower than the speed of light. For the 3+1 dimensional viscous fluid to be
causal in the shear channel, the following relation should be hold
\cite{dkkm3},
\begin{equation}
\frac{\eta}{\tau_{\pi}(\varepsilon+P)}\leq\frac{3}{4}(1-c_{s}^{2}),
\label{causalCD}%
\end{equation}
where $c_{s}$ is the ordinary sound velocity $c_{s}^{2}=dP/d\varepsilon$. In
particular, $c_{s}^{2}$ is $1/3$ in the massless ideal gas case and this
$\eta-\tau_{\pi}$ ration should be smaller than $1/2$.

Now we apply our formula to QCD related problems. In the hadronic phase at low
baryon density and high temperature, the dominant behavior comes from the
pions. The interaction of pions can well be described by the chiral
perturbation theory which is a low energy effective theory of QCD. To estimate
the GKN shear viscosity coefficient $\eta_{GKN}$, we adapt the leading order
result given in \cite{nakano}. At the high temperature regime far above the
deconfinement scale, we assume that the dominant behavior of the QGP is given
by gluon, and use $\eta_{GKN}$ obtained to the leading order in the
perturbative QCD (pQCD) calculation for $N_{f}=0$ in \cite{arnold}. We take
the running coupling constant to have the same parametrization as
\cite{cernai}.

In order to estimate $\eta$ and $\tau_{\pi}$ to the leading order in
couplings, it is enough to calculate the fluctuations of $J^{\mu}$ and
$\pi^{\mu\nu}$ to the lowest order. Then we obtain
\begin{align}
&  \int d^{3} \mathbf{x} (J^{x}(\mathbf{x}),J^{x}(\mathbf{0}))=\frac{g}{3}%
\int\frac{d^{3}p}{(2\pi)^{3}}\mathbf{p}^{2}n_{p}(1+n_{p})\nonumber\\
&  \hspace{3cm}=\frac{\varepsilon+P}{\beta},\label{leading_j}\\
&  \hspace{-1.5cm}\int d^{3} \mathbf{x} (\pi^{xy}(\mathbf{x}),\pi
^{xy}(\mathbf{0}))=\frac{g}{15}\int\frac{d^{3}p}{(2\pi)^{3}}\frac
{(\mathbf{p}^{2})^{2}}{\beta E_{p}^{2}}\left(  \frac{1}{E_{p}}+\beta
(1+n_{p})\right)  n_{p}\nonumber\\
&  \hspace{3cm}=\frac{P}{\beta}, \label{leading_pi}%
\end{align}
where $n_{p}=(e^{\beta E_{p}}-1)^{-1}$ is the Bose-Einstein distribution
function. The statistical factor $g$ is $3$ for pions and $16$ for gluons. In
the derivation of Eq.~(\ref{leading_pi}), we have subtracted the vacuum
contribution. Here $\varepsilon$ and $P$ are the energy density and pressure
of the ideal Bose gas, respectively.

By substituting them into the definitions Eqs.~(\ref{csvc}) and (\ref{relaxt}%
), we obtain the leading order results. In this approximation, we found that
the causal shear viscosity coefficient is given by $\eta=\eta_{GKN}$, and the
relaxation time (\ref{relaxt}) reduces to
\begin{align}
\tau_{\pi}=\eta_{GKN}/P=\eta/P.
\end{align}
See \cite{cernai,nakano,lacey} for the behavior of $\eta_{GKN}$.

\begin{figure}[ptb]
\includegraphics[scale=0.2]{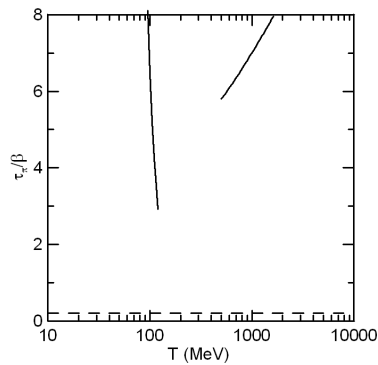}\caption{The temperature
dependence of the relaxation time to $\beta$ ratio with $\beta=1/T$. The solid
and dashed lines represent our result and that from AdS/CFT correspondence,
respectively. }%
\label{fig2}%
\end{figure}

We show the temperature dependence of $\tau_{\pi}$ to $\beta$ ratio in
Fig.~\ref{fig2}. The behavior is very similar to that of $\eta/s$: a
decreasing function of temperature in the hadron phase and an increasing one
in the QGP phase, exhibiting minimum near the phase transition temperature
$T_{c}$. The dashed line denotes the prediction from the AdS/CFT
correspondence $\tau_{\pi}/\beta=(2-\ln2)/(2\pi)$ for $\mathcal{N}=4$ SYM
\cite{baier}. However, because of the weak temperatrure dependence of
$\tau_{\pi}/\beta$ in the QGP phase, $\tau_{\pi}$ itself is a decreasing
function for both of the hadron and QGP phases and show discontinuity near
$T_{c}$, that is, $\tau_{\pi}$ becomes minimum (maximum) in the hadron (QGP)
side near $T_{c}$. Thus the behavior of the shear viscosity is insensitive for
the rapid change of the fluid velocity above $T_{c}$ because of the large
$\tau_{\pi}$, meanwhile it is more sensitive below $T_{c}$. In this sense, the
matter created in relativistic heavy-ion collisions can be close to the ideal
fluid in different two ways. one is because of the small $\eta$ and $\tau
_{\pi}$ which may be realized in the hadron phase and the other is the small
$\eta$ but large $\tau_{\pi}$ in the QGP phase.

\begin{figure}[ptb]
\includegraphics[scale=0.2]{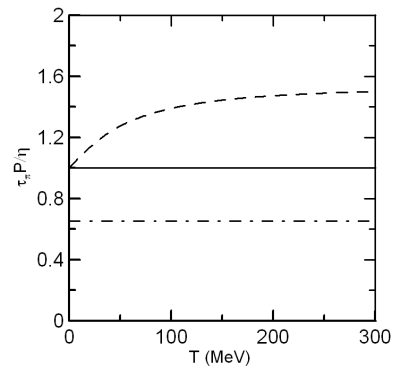}\caption{The temperature
dependence of $\tau_{\pi} P/\eta$. The solid, dashed and dot-dashed lines
correspond to the result of our formula, the moment method and AdS/CFT
correspondence, respectively.}%
\label{fig3}%
\end{figure}

The relaxation time is calculated also from the relativistic Boltzmann
equation using Grad's moment method with the 14 moments approximation
\cite{is,PPV}. In order to compare with our result, the temperature dependence
of $\tau_{\pi}$ to $\eta$ ratio in the unit of pressure $P$ is shown in Fig.
\ref{fig3}. The solid line denotes the behavior of our result, which should be
one in our leading order calculation. The dot-dashed line denotes the result
from the AdS/CFT correspondence, $(2-\ln2)/2$ \cite{baier}. The dashed line
represents the result obtained from the moment method for the pion
(Bose-Einstein) gas without phase transition \cite{is}. This figure shows that
our theory gives a different result from the "14" moments calculation. In the
Navier-Stokes limit (Newtonian fluid), it is known that the momentum method
with the "13" moments approximation and the famous Chapman-Enskog procedure
are consistent for the shear viscosity. However, $\tau_{\pi}$ requires
expansions in higher moments and, to see the relation of our formula and the
moment method, a more careful comparison should be done. It is also worth
mentioning that there is a attempt to calculate these coefficients without
using the moment method \cite{guy}.

As the AdS/CFT correspondence predicts that the minimum of $\eta_{GKN}/s$ has
a lower bound $1/(4\pi)$ \cite{son}, our $\eta/s$ and $\tau_{\pi}$
would have a lower bound, unless the fluctuation of $\pi^{\mu\nu}$ diverges.
As a matter of fact, to be consistent with the causality condition
(\ref{causalCD}), $\tau_{\pi}$ is somehow correlated with $\eta$, and
$\tau_{\pi}/\beta$ cannot be much smaller than $\eta/(\beta(\varepsilon+P))$.
\begin{figure}[ptb]
\includegraphics[scale=0.2]{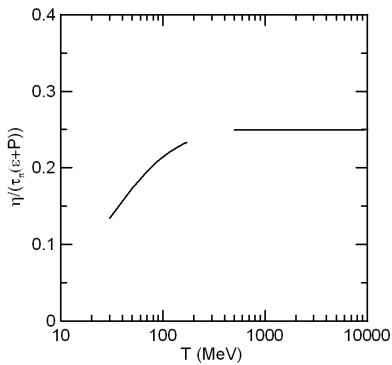}\caption{The temperature
dependence of the $\eta$-$\tau_{\pi}$ ratio. }%
\label{fig4}%
\end{figure}In Fig.~\ref{fig4} ,we show the temperature dependence of the
$\eta$-$\tau_{\pi}$ ratio. In our leading order calculaiton, this 
ratio is nothing but $\frac{\eta}{\tau_{\pi}(\varepsilon+P)}=\frac
{P}{(\varepsilon+P)}$, and shows a non-trivial temperature dependence
only at lower temperature than the pion mass and finally converges to the
massless ideal gas limit $1/4$ where $\varepsilon=3P$. One can easily see that
the result satisfies the relativistic causality condition (\ref{causalCD}). It
should be emphasized that relativistic fluids become unstable if the causality
condition is not satisfied \cite{dkkm3}. To see the consistency of CDH, 
it is necessary to investigate this ratio more carefully.

In summary, we proposed the compact definitions of the transport coefficients
of CDH, and calculated them in the chiral perturbation theory and
pQCD. We found that, in the leading order calculation, the causal shear
viscosity coefficient $\eta$ is reduced to that of the GKN shear viscosity
coefficient $\eta_{GKN}$. Although intuitive, this result is not trivial a
priori, because the irreversible current is not simply proportional to the
thermodynamic force in CDH. In fact, $\tau_{\pi}$ does not
vanish in our calculation, and there is a simple relationship, $\tau_{\pi
}=\eta/P$. This relation is not the same as that obtained in the 14 moment
calculation, and the physical origin of this discrepancy should be clarified.
So far, the improvements of the moment method has been discussed in diverse
ways \cite{grad_ce}. Our result may give a milestone in such a development.

It is thus very interesting to ask how these are modified when the calculation
is implemented beyond the leading order. The simple relationship obtained here
would not be satisfied in the strong coupling limit. In fact, from AdS/CFT
results, we can show $\eta/\tau_{\pi}=Ts/(4-2\ln2)$, which is larger than the
equation of state of the massless ideal fluid, $P=Ts/4$, that is, $P\neq
\eta/\tau_{\pi}$ as was shown in Fig.~\ref{fig3}. Thus we expect that the
simple relation $\eta=\eta_{GKN}$ and/or $\tau_{\pi}=\eta/P$ calculated from
the exact formulae (\ref{csvc}) and (\ref{relaxt}) would acquire a non-trivial
temperature dependence once going beyond the leading order result and in the
strong coupling limit.

T. Koide thanks to X. Huang and S. Pu for useful comments. This work was
(financially) supported by FAPERJ, CNPq and the Helmholtz International Center
for FAIR within the framework of the LOEWE program (Landesoffensive zur
Entwicklung Wissenschaftlich- Okonomischer Exzellenz) launched by the State of Hessen.

\end{document}